\documentclass[10pt,journal]{IEEEtran}
\makeatletter
\def\ps@headings{%
\def\@oddhead{\mbox{}\scriptsize\rightmark \hfil \thepage}%
\def\@evenhead{\scriptsize\thepage \hfil \leftmark\mbox{}}%
\def\@oddfoot{}%
\def\@evenfoot{}}
\makeatother

\usepackage{amsmath}
\usepackage{amssymb}
\usepackage{bbm}
\usepackage{graphics}
\usepackage{psfrag}
\usepackage{epsfig}
\usepackage[verbose]{cite}
\usepackage{algorithm}
\usepackage{algorithmic}
\usepackage{cite}
\newcommand{\ora}{\overrightarrow}

\newcommand{\E}{\ora{E}}

\newcommand{\ve}{\varepsilon}

\newcommand{\vp}{\varphi}

\newcommand{\mce}{\mathcal{E}}

\newcommand{\mcm}{\mathcal{M}}
\newcommand{\mcr}{\mathbf{r}}
\newcommand{\mcp}{\mathbf{p_b}}
\newcommand{\bfn}{\mathbf{n}}

\newcommand{\mcc}{\mathcal{C}}

\newcommand{\ol}{\overline}

\newcommand{\mbb}{\mathbbm}

\newtheorem{thm}{Theorem}
\newtheorem{cor}{Corollary}

\title{Study of Throughput and Delay in Finite-Buffer Line Networks}
\author{
   \small
   \begin{tabular}{p{.5\textwidth} p{.5\textwidth}}
        \hspace{15mm}
        \begin{minipage}{.4\textwidth}
         \centering
         {Badri N. Vellambi} \\
         Institute for Telecommunications Research\\
         University of South Australia \\
         Mawson Lakes, Australia \\
         E-mail: \{badri.vellambi\}@unisa.edu.au
      \end{minipage}&    
      \begin{minipage}{.4\textwidth}
         \centering
         {Nima Torabkhani, Faramarz Fekri} \\
         School of Electrical and Computer Engineering \\
         Georgia Institute of Technology \\
         Atlanta, GA 30332-0250 \\
         E-mail: \{nima, fekri\}@ece.gatech.edu
      \end{minipage}           
   \end{tabular} \vspace{-8mm}}

\begin{document}
\maketitle
\begin{abstract}
In this work, we study the effects of finite buffers on the throughput and delay of line networks with erasure links. We identify the calculation of performance parameters such as throughput and delay to be equivalent to determining the stationary distribution of an irreducible Markov chain. We note that the number of states in the Markov chain grows exponentially in the size of the buffers with the exponent scaling linearly with the number of hops in a line network. We then propose a simplified iterative scheme to approximately identify the steady-state distribution of the chain by decoupling the chain to smaller chains. The approximate solution is then used to understand the effect of buffer sizes on throughput and distribution of packet delay. Further, we classify nodes based on congestion that yields an intelligent scheme for memory allocation using the proposed framework. Finally, by simulations we confirm that our framework yields an accurate prediction of the variation of the throughput and delay distribution.
\end{abstract}

\vspace{-2mm}
\section{Introduction}\label{FB-intro}

In networks, packets that have to be routed from one node to the other may have to be relayed through a series of intermediate nodes. Also, each node in the network may receive packets via many data streams that are being routed simultaneously from their source nodes to their respective destinations. In such conditions, the packets may have to be stored at intermediate nodes for transmission at a later time. If buffers are unlimited, the intermediate nodes need not have to reject or drop packets that arrive. For practical reasons, buffers are limited in size. Although a large buffer size is preferred to minimize packet drops, large buffers have an adverse effect on the mean and variance in packet delay. Additionally, as second-order effects, using larger buffer sizes at intermediate nodes would have practical problems such as on-chip board space and increased memory-access latency. Though our work is motivated partly by such concerns, our work is far from modeling realistic scenarios. This work modestly aims at providing a theoretical framework to understand the fundamental limits of single information flow in finite-buffer line networks and investigate the trade-offs between throughput, packet delay and buffer size.

The problem of computing capacity\footnote{Throughout this work, we use capacity to mean the supremum of all rates of information flow achievable by any coding scheme.} and designing efficient coding schemes for lossy wired and wireless networks has been widely studied \cite{AmirDana:capacity, PakzadF05, NCAlgApproch, YeungNCJrnl}. However, the study of capacity of networks with finite buffer sizes has been limited. This can be attributed solely to the fact that the analysis of finite buffer systems are generally more challenging. With the advent of network coding \cite{YeungNCJrnl,LUN04} as an elegant and effective tool for attaining optimum network performance, the interest in the study of finite buffer networks has been increased.

The problem of studying throughput and delay of networks with finite buffers has also been studied in queueing theory. These problems can be seen to be similar,  since the packets can be viewed as customers and the delay due to packet loss in the link as the arbitrary service time. Also, the phenomenon of packet overflow in the network can be modeled by a type II blocking (commonly known as \emph{blocking after service}) in stochastic networks. However, there is a subtle difference in the packet-customer analogy when the network has nodes that can send packets over multiple paths to the destination. When such is the case, the node can choose to duplicate packets on both the paths, an event that cannot be captured directly in the customer-server based queueing model. However, that is not the case in line networks. Therefore, the problem of finding buffer occupancy distribution and consequently throughput and delay in certain networks is then seen to be identical to determining certain arrival/departure processes in an open stochastic network of a given topology \cite{Tayfur_QT1, Brandwajn_QT3, SO_QT4, Serfozo_QT0}. Such relevant works in the field of queueing theory consider a continuous-time model for arrival and departure of packets in the network.

In~\cite{NetCod:Lun}, Lun \emph{et al.} consider the discrete-time analogue of the arrival process by lumping time into epochs (wherein each node can transmit and receive a packet) to analyze the capacity of a simple two-hop lossy network. In our previous work~\cite{VelambiITW}, we derived bounds on the throughput of line networks, which were unable to provide good approximations for packet delay and buffer occupancy statistics. While our approach employs a model of network similar to that in \cite{NetCod:Lun, VelambiITW}, we extend their results not only to derive estimates for the capacity of line networks of any hop-length and intermediate node buffer size, but also to derive quantitative estimates for packet delay distribution. Our contributions to this area of research is summarized below.
\begin{itemize}
\item[1.] We extend a Markov-chain based modeling to present an iterative estimate for the buffer occupancy distribution at intermediate nodes.
\item[2.] Using the estimate, we derive expressions for throughput and packet delay distribution that are seen to be fairly accurate in predicting the actual system behavior.
\end{itemize}

This work is organized as follows. First, we present the formal definition of the problem and the network model in
Section~\ref{FB-Model}. Next, we introduce our analysis for finite-buffer line networks in Section~\ref{FB-Analysis} and then investigate packet delay in Section~\ref{FB-Delay}. We compare our analytical results with actual simulations in Section~\ref{FB-Results} and in Section~\ref{FB-Discussion} we briefly discuss trade-offs between throughput, delay and memory. Finally, Section~\ref{Conc} concludes the paper.

\section{Problem Statement and Network Model}\label{FB-Model}

\begin{figure}[h]
\psfrag{e0}{$\ve_1$} \psfrag{e1}{$\ve_2$} \psfrag{e2}{$\ve_h$} \psfrag{m1}{\hspace{-2mm}$m_1$} \psfrag{m2}{\hspace{-2mm}$m_2$} \psfrag{m3}{\hspace{-5mm}$m_{h-1}$}
\psfrag{v0}{$v_0$} \psfrag{v1}{$v_1$} \psfrag{v2}{$v_2$} \psfrag{v3}{\hspace{-2mm}$v_{h-1}$} \psfrag{v4}{$v_h$}
\centering
\includegraphics[width=2.5in, height=0.65in,angle=0]{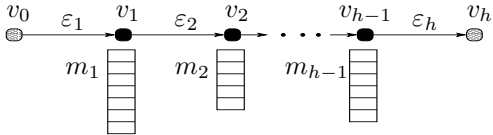}
\vspace{-2mm}
\caption{An illustration of the line network.}
 \label{LineNet}
\end{figure}

As illustrated in Figure~\ref{LineNet}, a line network is a directed graph of $h$ hops with
$V=\{v_0,v_1,\ldots,v_h\}$ and $\E=\{(v_i,v_{i+1}): i=0,\ldots,h-1\}$ for some integer $h\geq 2$. In the figure, the intermediate nodes are shown by black ovals. The links are assumed to be unidirectional, memoryless and lossy. We let $\ve_i$ denote the packet erasure probability over the link $(v_i,v_{i+1})$. The erasures model only the quality of links (e.g., presence of noise, interference) and do not represent packet drops due to finite buffers. A lossless hop-by-hop acknowledgement setup is in place to indicate the successful receipt of a packet\footnote{This assumption is made to simplify modeling. In the absence of perfect ACK, one can use random linear coding over a large finite field to achieve the same desired throughput.}. Moreover, the packet processes on different links are assumed to be independent. Each node $v_i\in V$ has a buffer of $m_i$ packets with each packet having a fixed size of $S$ bytes. Note that the buffer size can vary with the node index. Lastly, the source and destination nodes are assumed to have sufficient memory to store any amount of data.

The system is analyzed using a discrete-time model, where each node can transmit at most one packet over a link per epoch. We let $\{X_i(l)\}_{\mbb{Z}_{\geq 0}}$ to be the random process denoting the erasure occurrences on the link
$(v_{i-1},v_i)$ at time $l$. We set $X_i(l)=0$ if a packet is erased on $(v_{i-1},v_{i})$ at epoch $l$ and $X_i(l)=1$ otherwise.

The unicast capacity between a pair of nodes is defined to be the supremum of all achievable rates of transmission of
information packets (in packets per epoch) between a pair of nodes. The supremum is calculated over all possible means
used for packet generation and buffer update at intermediate nodes. Note that the source node can generate innovative packets during each epoch.
For instance, in the particular case of the line network of Figure~\ref{LineNet}, we would like to identify the unicast capacity between $v_0$ and $v_h$.

\begin{figure}[h]
\vspace{-3mm}
\centering
\psfrag{data2}{\scriptsize{Discrete-Time ($0.1$ Sec)}}
\psfrag{data3}{\scriptsize{Discrete-Time ($0.025$ Sec)}}
\psfrag{data1}{\scriptsize{Continuous-Time}}
\psfrag{ddddddddddddddddddddddddata4}{\scriptsize{Discrete-Time ($0.001$ Sec)}}
\psfrag{xaxis}{\hspace{-10mm}\footnotesize{Buffer size (in packets)}}
\psfrag{yaxis}{\hspace{-15mm}\footnotesize{Capacity (in packets/epoch)}}
\includegraphics[width=3.5in, height=1.75in,angle=0]{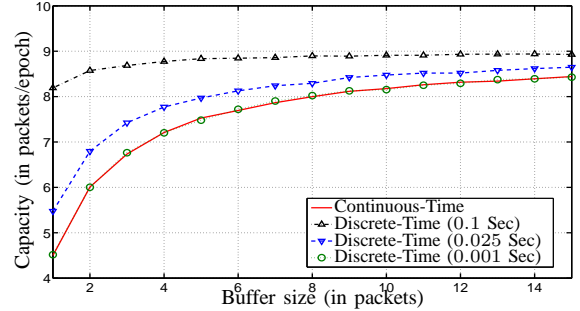}
\vspace{-8mm}
\caption{An illustration of the precision of the discrete-time model.}
 \label{Disc_Cont}
\end{figure}

 Before we proceed to the modeling, we briefly motivate the assumed discrete-time model with an example. Consider a continuous-time model with the discrete-time model for varying times of epoch for a simple continuous-time two-hop line network with a Poisson packet generation process at the source with parameter $\lambda_1=10$ pkts/sec. The service time at the intermediate node is also Poisson with parameter $\lambda_2=10$ pkts/sec, and the links connecting the source to the intermediate node and the intermediate node to the destination are both packet-erasure channels with erasure probabilities $\ve_1={\ve_2}=0.1$. Finally, suppose that the intermediate node has a finite buffer of $m=10$ packets. Figure~\ref{Disc_Cont} presents the (simulated) capacity for the continuous model and the time-discretized models for various epoch durations. It is noticed that as the epoch duration is made smaller, the discrete-time model becomes more accurate in predicting the capacity. This was verified to be the case for all line networks with Poisson arrivals and service times.

%

Lastly, we use the following notations. $\mathbbm{G}(t)$ denotes the geometric distribution with mean inter-arrival time $t\in\mathbbm{R}$. $\sigma(\cdot)$ denotes the indicator function for $\mathbbm{Z}_{> 0}$. For any $x\in\mathbbm{R}$, $\ol x \triangleq 1-x$. Lastly, $\otimes$ denotes the convolution operator.

\vspace{-.025 in}
\section{Finite-Buffer Analysis}
\label{FB-Analysis}
\vspace{-.025 in} 
Here, we develop our framework for analyzing finite-buffer erasure networks. In the sections that follow, these techniques will be used to study various the performance indicators. We proceed as follows.
\subsubsection{Rate-optimal Schemes}
\label{RateOptimality}
One of the most important performance parameters of a network is its throughput and the problem of identifying capacity is directly related to the problem of finding schemes that are \emph{rate-optimal}. In our model of line network, a scheme that performs the following in the same order can be seen to be rate-optimal.
\begin{itemize}
\item[1.] If the buffer of a node is not empty at a particular epoch, then it must transmit at least one of the packets.
\item[2.] A node deletes the packet transmitted at an epoch if it receives an acknowledgement from the next hop.
\item[3.] A node accepts an arriving packet if it has space in its buffer. It then sends an ACK to the previous node.
\end{itemize}
In the absence of feedback, rate-optimality can be achieved by employing random linear combinations based network coding over a large finite field as is described in~\cite{VelambiITW, NetCod:Lun}.

\subsubsection{Markov Chain Modeling of the Buffer States at Intermediate Nodes}
\label{sec-MC-SR}
In order to model the network with lossless feedback, we need to track the number of packets that each node possesses at
every instant of time. We do so by using the rules of buffer update under the optimal scheme. Let
$\mathbf{n}(l)=(n_1(l),\ldots,n_{h-1}(l))$ be the vector whose $i^{\textrm{th}}$ component denotes the number of packets the
$i^{\textrm{th}}$ intermediate node possesses at time $l$. The variation of state at the $l^{\textrm{th}}$ can be tracked
using auxiliary random variables $Y_i(l)$ defined by

\vspace{-3mm}
\begin{equation*}
\small
Y_i(l)\hspace{-0.5mm}=\hspace{-0.5mm}\left\{\hspace{-1.5mm}\begin{array}{ll}\sigma(n_{i-1}(l))X_i(l) & i=h\\X_i(l)\sigma(m_{i}-n_{i}(l)+Y_{i+1}(l)) & i=1\\
\sigma(n_{i-1}(l))X_i(l)\sigma(m_{i}-n_{i}(l)+Y_{i+1}(l)) & 1<i<h
 \end{array}\right.\label{FB-eqn2}
\end{equation*}

The above definition is such that $Y_i(l)=1$ only if all the
following conditions are met.
\begin{itemize}
\item[1.] Node $v_{i-1}$ has a packet to transmit to $v_{i}$.
\item[2.] The link $(v_{i-1},v_i)$ does not erase the packet at the $l^{\textrm{th}}$ epoch, \emph{i.e.}, $X_{i}(l)=1$.
\item[3.] Node $v_{i}$ has space after it has updated its buffer for any changes due to its transmission at that epoch.
\end{itemize}
The dynamics of buffer states can then by seen to be
\begin{equation}
n_i(l+1)=n_i(l)+Y_i(l)-Y_{i+1}(l),\quad\quad i=1,\ldots,h-1.\label{FB-eqn3}
\end{equation}
Note that since $\mathbf{Y}(l)=(Y_1(l),\ldots,Y_h(l))$ is a function of $\bfn(l)$ and $\mathbf{X}(l)=(X_1(l),\ldots,X_h(l))$, $\bfn(l+1)$ depends only on its previous state $\bfn(l)$ and the channel conditions $\mathbf{X}(l)$ at the
$l^{\textrm{th}}$ epoch. Hence, we see that $\{\bfn(l)\}_{l\in\mathbbm{Z}_{\geq 0}}$ forms a Markov chain. It is readily checked that this chain has $\prod_{i=1}^{h-1} (m_i+1)$ states. Further, this chain is \emph{irreducible, aperiodic, positive-recurrent}, and \emph{ergodic}~\cite{MCBook} and therefore has a unique steady-state probability $p^\infty$. By ergodicity, we can obtain temporal averages by statistical averages. We then see that the computation of (throughput) capacity is equivalent to the computation of the likelihood of the event that $Y_h=1$.

\subsubsection{Approximated MC for an intermediate node}
\label{Appx-MC}
The exponential growth in the size of the chain and the presence of boundaries (due to finite buffers), exact calculation of the steady-state probabilities (and hence the throughput) becomes very cumbersome even for networks of reasonable buffer sizes and hop-lengths. The exact chain for the dynamics of the system is such that a state update at a node has a strong dependence on the states of both its previous-hop and its next-hop neighbors. Additionally, the process of packet transmission over intermediate edge can be shown to be non-memoryless. These facts add to the intractability of the exact computation of the distribution. However, it is possible to decouple the chain into several Markov chains with a single finite-boundary under some simplifying assumptions. To have an approximate decoupled model, we need to identify the transition probabilities of the decoupled chains, which is possible only if we know the arrival and departure processes on each edge. The rate of information on any edge is directly related to the fraction of time the sending node is non-empty and the fraction of time a successfully delivered packet will get blocked (and this happens if the receiving node is full at the time of packet arrival). Hence, to have a model for a node, we need to have the approximate buffer occupancy distributions for neighboring nodes. This hints naturally at an \emph{iterative} approach to the problem. In this section, we develop an iterative estimation method that considers the effect of blocking with some simplifying assumptions. To develop an iterative technique, we assume the following.
\begin{itemize}
\item[A1.] The packets are ejected from nodes in a memoryless fashion. Equivalently, we assume that $\Pr[(n_{i-1}(t)>0)\wedge
(X_i(t)=1) | n_i(t)=k ]$ does not vary with the occupancy $k$ of the $i^{\textrm{th}}$ node. This allows us
to track just the information rate and not the exact statistics.
\item[A2.] The blocking event occurs independent of the state of a node, \emph{i.e.},
$\Pr[(Y_{i+1}(l)=0)\wedge(X_{i+1}(l)=1)|(n_i(t)=k)]$ is the same for $k=1,\ldots, m_i$. This allows us to track just the
blocking probability and not the joint statistics.
\item[A3.] At any epoch, given the occupancy of a particular node, the arrival process is independent of the blocking process.
\end{itemize}
\begin{figure}[ht!]
\psfrag{aa}{$\scriptsize{\ol\alpha_0}$}
\psfrag{bb}{$\scriptsize{\hspace{-4mm}\ol{\alpha}-\beta}$}
\psfrag{cc}{$\scriptsize{\ol\beta}$}
\psfrag{dd}{$\scriptsize{\alpha_0}$}
\psfrag{ee}{$\scriptsize{\alpha}$}
\psfrag{ff}{$\scriptsize{\beta}$}
\psfrag{gg}{$\scriptsize{\beta}$}
\psfrag{S0}{\scriptsize{0}}
\psfrag{S1}{\hspace{1mm}\scriptsize{1}}
\psfrag{S2}{\small{2}}
\psfrag{S3}{\hspace{-2mm}\scriptsize{${m_i}$\hspace{-0.25mm}-\hspace{-0.25mm}1}}
\psfrag{S4}{$\hspace{-2mm}\scriptsize{{m_i}}$}
\centering
\includegraphics[height=.8in,width=3.35in,angle=0]{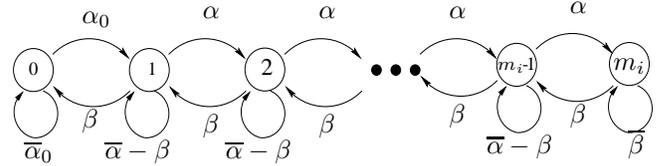}\vspace{-3mm}
\caption{The chain for the node $v_i$ obtained by the assumptions A1-A3.}
\label{FB-1HMC}
\end{figure}
These assumptions spread the effect of blocking equally over all non-zero states of occupancy at each node. Under these assumptions, when we are given that the arrival rate of packets as seen by $v_i$ is $r_i$ packets/epoch, and that the probability of blocking by $v_{i+1}$ is
${p_b}_{i+1}$, the dynamics of the state change for the node $v_i$ is given by the chain in Figure~\ref{FB-1HMC} with the parameters set to
\begin{equation}
\small{
\begin{array}{ccl}
\alpha&=&r_i(\ve_{i+1}+\ol\ve_{i+1}{p_b}_{i+1})\\
\beta&=&(1-r_i)\ol{p_b}_{i+1}\ol\ve_{i+1}\\
\alpha_0&=&r_i
\end{array}}.
\end{equation}
For this chain in Fig.~\ref{FB-1HMC}, the steady-state distribution can be computed to be
\begin{equation*}
{
\tiny
\Pr[n_i=k]=\vp(k| r_i,\ve_{i+1},{p_b}_{i+1})\triangleq\left\{\begin{array}{cc}\frac{1}{1+\frac{\alpha_0}{\beta}\big(\sum_{l=0}^{m_i-1}\frac{\alpha^l}{\beta^l}\big)} & k=0\\   \\
\frac{\frac{\alpha_0\alpha^{k-1}}{\beta^k}}{1+\frac{\alpha_0}{\beta}\big(\sum_{l=0}^{m_i-1}\frac{\alpha^l}{\beta^l}\big)} & k\neq 0\end{array}\right. \label{FB-SS}}
\end{equation*}
The blocking probability that the node $v_{i-1}$ perceives from the node $v_i$, assuming that $v_i$ sees an arrival rate of
$r_i$ from $v_{i-1}$ and a blocking probability of ${p_b}_{i+1}$ caused by $v_{i+1}$\footnote{Note that the arrival rate
at the node $v_1$ is $r_1=\ol\ve_1$ and that the blocking probability of $v_h$ is zero.}, can then be calculated as follows.
\begin{equation}
\hspace{-2mm}{
\tiny
p_b}_i\hspace{-1mm}=\hspace{-1mm}\left\{\hspace{-1.5mm}\begin{array}{ll} (\ve_{i+1}\hspace{-0.5mm}+\hspace{-0.5mm}\ol\ve_{i+1}{p_b}_{i+1})\vp(m_i| r_i,\ve_{i+1},{p_b}_{i+1}) & \hspace{-2mm}i\hspace{-0.5mm}<\hspace{-0.5mm}h\hspace{-0.5mm}-\hspace{-0.5mm}1 \\
 \ve_{i+1}\vp(m_i| r_i,\ve_{i+1},0) &\hspace{-2mm} i\hspace{-0.5mm}=\hspace{-0.5mm}h\hspace{-0.5mm}-\hspace{-0.5mm}1 \end{array}\right. \label{FB-PB}
\end{equation}
Similarly, the arrival rate at on each edge can be seen to be related to the occupancy of the previous node and the channel erasure probability in the following fashion.
\begin{equation}
r_{i+1}=\left\{\begin{array}{ll}\ol\ve_i (1-\vp(0| r_i,\ve_{i+1},{p_b}_{i+1})) &\hspace{-1mm} 1<i<h-1 \\
\ol\ve_i (1-\vp(0| r_i,\ve_{i+1},0)) &\hspace{-1mm} i=h-1 \end{array}\right. \label{FB-R}
\end{equation}

Given two vectors $\mcr=(r_1,\ldots,r_h)\in[0,1]^{h}$ and $\mcp=({p_b}_1,\ldots,{p_b}_h)\in[0,1]^{h}$, we term $(\mcr,\mcp)$
as an approximate solution to the chain, if they satisfy the equations (\ref{FB-PB}), and (\ref{FB-R}) in addition to having $r_1=\ol\ve_1$ and ${p_b}_h=0$. The following theorem guarantees both the uniqueness and the method of
identifying the approximate solution to the chain.
\begin{thm}\label{FB-uniqthm}
Given a line network with link erasures $\mce=(\ve_1,\ldots,\ve_h)$ and intermediate node buffer sizes $\mcm=(m_1,\ldots,
m_{h-1})$, there is exactly one approximate solution $(\mcr^*(\mce,\mcm),\mcp^*(\mce,\mcm))$ to the chain. Moreover, this
solution can be found iteratively.
\end{thm}
While the proof of uniqueness is omitted for brevity, the method of identifying the approximate solution follows a simple iterative scheme. First, $\mcp$ is set to $\mathbf{0}$ and \ref{FB-R} is invoked to estimate $\mcr$. This estimate is then used in \ref{FB-PB} to update $\mcp$, which is then used to find the next estimate of $\mcr$. These iterations repeated until the vectors converge. We omit the details of proof of convergence for lack of space.

Finally, the unicast capacity $\mcc^*(\mce,\mcm)$ can be estimated using the approximate solution from the following.
\begin{cor}
The approx. solution $(\mcr^*(\mce,\mcm),\mcp^*(\mce,\mcm))$  satisfies the flow conservation relations

\hspace{-5mm} $\quad$ $\mcc^*(\mce,\mcm)=r^*_1(1-{p_b}^*_1)=r^*_i(1-{p_b}^*_i), \,\,i=1,\ldots,h. $
\end{cor}
\vspace{-1mm}
\section{Packet Delay Distribution}
\label{FB-Delay}

In this section, we use the approximate solution of Section~\ref{Appx-MC} to obtain the estimates on the probability distribution of the delay of an information packet. We define the packet delay as the time taken from the instant when the source starts sending the packet to the instant when the destination receives it. In addition to the discussion in section \ref{RateOptimality}, we assume a \emph{first-come first-serve} treatment of packets at the intermediate node buffers.

In order to compute the distribution of delay that a packet experiences in the network, one can proceed in a hop-by-hop fashion. Considering the last relay node, the additional delay of an arriving packet (at time $l$) at node $v_{h-1}$ depends on the occupancy of the node $v_{h-1}$ and the erasure channel that follows it to the destination. Suppose at epoch $l$, node $v_{h-1}$ has $k\leq m_{h-1}-1$ packets in addition to the arriving packet. Then, the packet has to wait for the first $k$ packets to leave before it can be serviced. Since each transmission takes place independently, the distribution of delay is sum of $k+1$ independent geometric distribution with mean inter-arrival time $\frac{1}{1-\ve_h}$, which is denoted by ${\otimes^{k+1}\mathbbm{G}(\frac{1}{1-\ve_h})}$. Suppose that the distribution of buffer occupancy \emph{at time of packet arrival} is given by $\pi_{h-1} (i)$, then the distribution of delay added by $v_{h-1}$ to the packet is
\begin{equation}
\small{\mathbf{D}_{h-1}=\sum_{i=0}^{m_{h-1}-1} {\pi_{h-1} (i) {\otimes^{i+1}\mathbbm{G}((1-\acute{\ve_h})^{-1})}}.     \label{last_relay_delay}}
\end{equation}
However, the situation is different for other intermediate delays because of the effect of blocking. The additional delay incurred while being stored at the node $v_j,\, 0<j<h-1$, is given by
\begin{equation}
\small{\mathbf{D}_{j}=\sum_{i=0}^{m_{j}-1} {\pi_{j} (i) {\otimes^{i+1}\mathbbm{G}((1-\acute{\ve_{j+1}})^{-1})}},     \label{int_relay_delay}}
\end{equation}
where we used the following to consider blocking.
\begin{equation}
\small{{\ve'_i}=\left\{\begin{array}{ll} \ve_i + \theta_{v_i}(m_i) (1-\ve_i) & i=1,2,\ldots,h-1\\\ve_h & i=h
\end{array}\right.,}
\end{equation}
where $\theta_{v_i}(k)$ is the steady state probability of that node $v_i$ already has $k$ packets when the packet is transmitted successfully from $v_{i-1}$. $\pi_j (i)$ and $\theta_{v_i}(k)$ are related by
\begin{equation}
\small{\pi_j (i)=\left\{\begin{array}{ll} \frac{\theta_{v_j}(i)}{1-\theta_{v_j}(m_j)} & i=1,2,\ldots,m_j-1\\ 0 & i=m_j
\end{array}\right..}
\end{equation}
By assuming that the delays incurred by each node and its adjoining outgoing link is independent of each other, we obtain the total delay considering all hops to be
\begin{equation}
\small{\mathbf{D}=\mathbbm{G}((1-\acute{\ve_1})^{-1})\otimes \mathbf{D}_{1}\otimes \cdots \otimes \mathbf{D}_{h-1}.}
\label{Delay_dist_formula}
\end{equation}

Hence, the delay distribution is known if the steady-state distributions of buffer states ($\pi_j(\cdot),\,\, j=1,...,h-1$) as seen by arriving packets is known. However, it is a simple exercise to derive these distributions from the results of Section \ref{Appx-MC}.

%
%
%
%
%
\vspace{-1mm}
\section{Results on Finite-Buffer Analysis}
\label{FB-Results}
\vspace{-.025 in} 

We have so far presented some fundamental tools for finite-buffer analysis of line networks. In this section, we show that they are very helpful to obtain accurate estimates of the performance parameters such as throughput, delay distribution and buffer occupancy distribution for line networks.

\begin{figure}[ht!]
\centering
\psfrag{xaxis}{\footnotesize{\hspace{7mm} Number of hops $h$}}
\psfrag{yaxis}{\small{\hspace{-5mm} Capacity (packets/epoch)}}
\psfrag{ddddddddddddddddddddata1}{\tiny{Sim. Cap. ($\ve=0.25$)}}
\psfrag{data2}{\tiny{Sim. Cap. ($\ve=0.50$)}}
\psfrag{data3}{\tiny{Iter. Estm.  ($\ve=0.50$)}}
\psfrag{data4}{\tiny{Iter. Estm.  ($\ve=0.50$)}}
\includegraphics[height=1.4in,width=3.5in,angle=0]{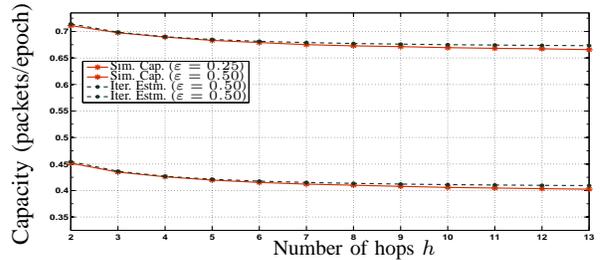}\vspace{-4mm}
\caption{Capacity of a line network with $m=5$ vs. the number of hops $h$.}\label{FB-LineL}
\end{figure}

To understand the variation of our capacity estimate of Section~\ref{FB-Analysis}, in each of the figures, the simulation of the actual capacity is presented in addition to our analytical results. Figure~\ref{FB-LineL} presents the variation of the capacity
with the hop length for a network with each intermediate node having a buffer size of five packets. Moreover, the simulations are performed when the probability of erasure on every link is set to either $0.25$ or $0.5$. It is noticed that the estimate captures the variation of the actual capacity of the network within about $1.5\%$ of error.

\begin{figure}[ht!]
\centering
\psfrag{xaxis}{\footnotesize{\hspace{10mm}Buffer size $m$}}
\psfrag{yaxis}{\small{\hspace{-4mm}Capacity (packets/epoch)}}
\psfrag{ddddddddddddddddddddata1}{\tiny{Sim. Cap. ($\ve=0.25$)}}
\psfrag{data2}{\tiny{Sim. Cap. ($\ve=0.50$)}}
\psfrag{data3}{\tiny{Iter. Estm.  ($\ve=0.50$)}}
\psfrag{data4}{\tiny{Iter. Estm.  ($\ve=0.50$)}}
\includegraphics[height=1.4in,width=3in,angle=0]{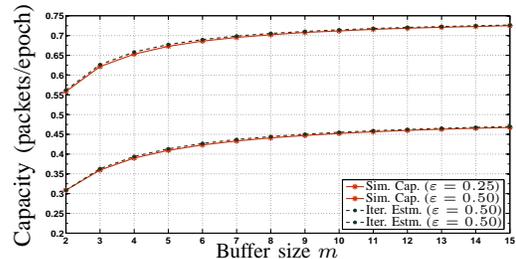}\vspace{-4mm}
\caption{Capacity of a line network with $h=8$ vs. the buffer size $m$.}\label{FB-LineM}\vspace{-3mm}
\end{figure}
In order to study the effect of buffer size, we simulated a line network
of eight hops having the same erasures as in the previous setting. Figure~\ref{FB-LineM} presents the variation of our results and the
actual capacity as the buffer size of the intermediate node is varied. It can be seen that as the buffer size is increased,
all curves approach the ideal min-cut capacity of $1-\ve$.

\begin{figure}[ht!]
\centering
\vspace{-1mm}
\includegraphics[height=2in,width=3.75in,angle=0]{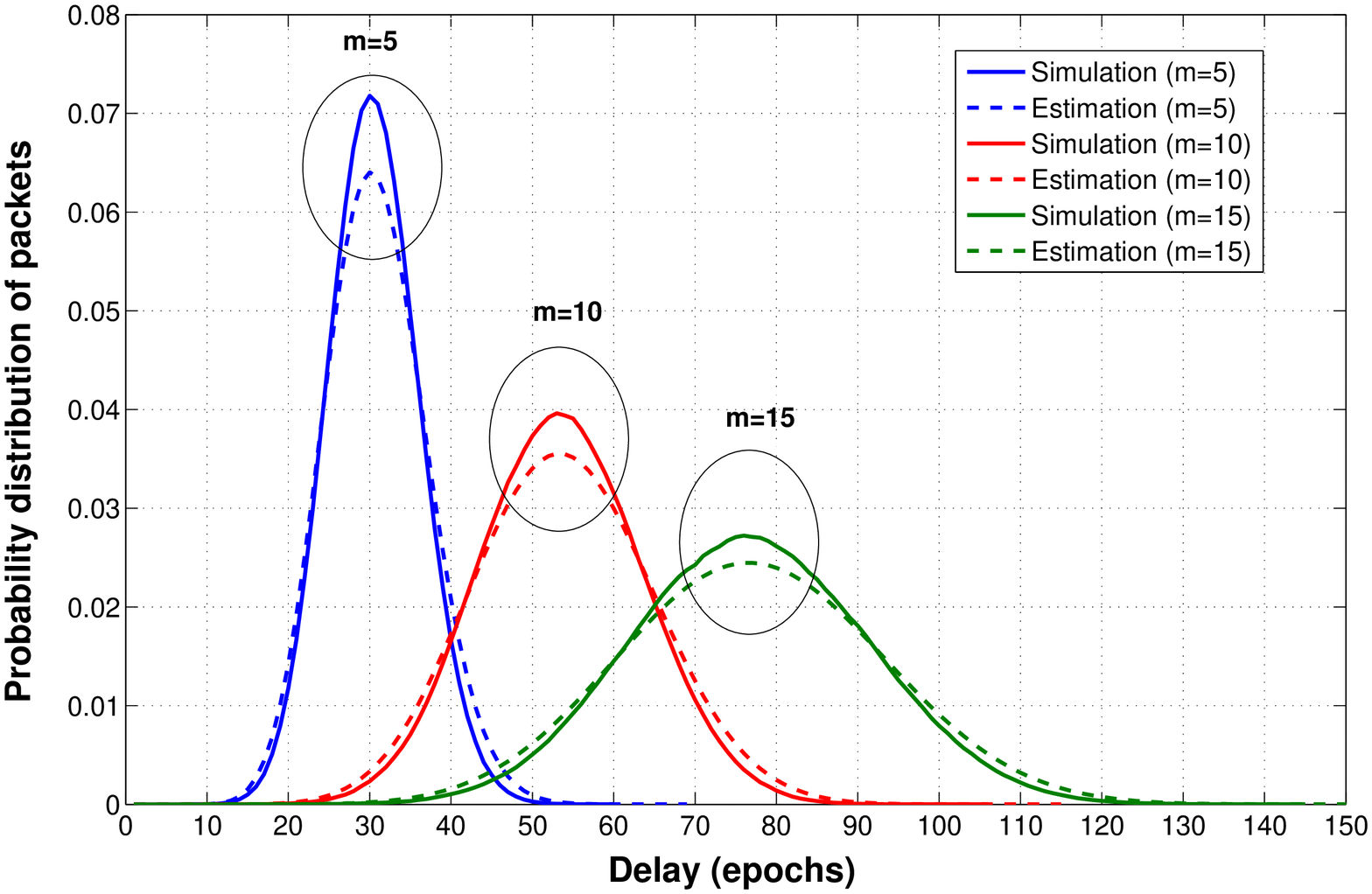}\vspace{-4mm}
\caption{Delay distribution in an 8 hop line network for varying buffer sizes.}\label{Delaydist-Line}
\vspace{-1mm}
\end{figure}

 Figure~\ref{Delaydist-Line} presents the variation of delay distribution with respect to the buffer size for an eight-hop line network with the erasure probability on every link set to $0.25$. It can be seen that both the mean and the variance of the distribution increases with the increase in the buffer size. It is noted that the analytic prediction of the delay is more conservative than the actual simulation i.e., the analytic estimate of the variance is higher than the actual simulated one.

\vspace{-1mm}
\section{Throughput and Delay trade-offs}
\label{FB-Discussion}
\vspace{-.01 in} 

\begin{figure}[htbp!]
\centering
\vspace{-1mm}
\includegraphics[height=2in, width=3.5in,angle=0]{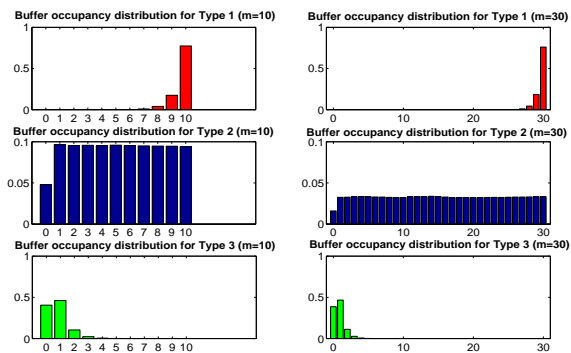}        \vspace{-10mm}
\caption{Buffer occupancy distribution for different types of nodes}
\label{Buffer-Occupancy}
\end{figure}

Based on our the computed buffer occupancy distributions, we categorize the nodes into 3 types according to their buffer occupancy. Nodes are of Type 1, 2 or 3 depending on whether the rate of incoming rate of innovative packets is larger, same, or smaller, respectively, than the  possible rate of innovative packets that can be sent on the outgoing edge. The typical occupancy distributions of such nodes are presented in Figure~\ref{Buffer-Occupancy}. The figure presents a classification of nodes from a four-hop line network with $\varepsilon_1=0.2, \varepsilon_2=0.5, \varepsilon_3=0.5, \varepsilon_4=0.2$ when the three intermediate nodes have buffer sizes $m=10$ and $m=30$, respectively. Note that in this example, Node $v_i$ is of Type $i$. While Type 3 nodes are generally starved and Type 1 nodes are generally full, Type 2 nodes have a near uniform distribution. Increasing the buffers of Type 1 or Type 3 nodes does not affect their blocking probabilities or the general shape of the occupancy distribution. However, increasing the buffer sizes of Type 2 nodes decreases the blocking probability of such nodes. Note that while the classification of these nodes is a trivial exercise, the means to identify/estimate the arrival and departure rates is non-trivial.

\begin{figure}[htbp!]
\centering
\vspace{-1mm}
\psfrag{Node 1}{\scriptsize{Type 1}}
\psfrag{Node 2}{\scriptsize{Type 2}}
\psfrag{Node 3}{\scriptsize{Type 3}}
\includegraphics[height=2.15in, width=3.5in, angle=0]{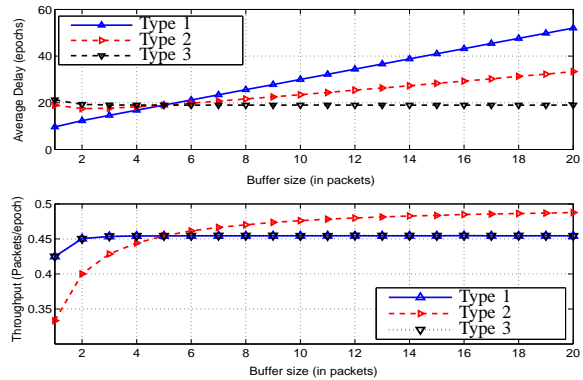}    \vspace{-8mm}
\caption{Throughput Vs. Average Delay for different type of nodes}
\label{Delay-Throughput}
\end{figure}

Figure~\ref{Delay-Throughput} shows the effect of increasing buffer size of nodes of a particular type on the throughput and delay in the mentioned line network example. In order to observe the effect of the chosen node of a particular type, the buffers of nodes of other types are fixed at five packets and buffer size of the desired node is varied. We conclude that by allocating memory to Type 1 nodes yields minimal improvement in throughput and a large increase in delay. However, the effect of increasing the buffers of Type 3 nodes has almost no effect on both parameters. However, allocating memory to Type 2 nodes can affect the throughput sufficiently although with a moderate increase in the expected delay.

\vspace{-3mm}
\section{Conclusions}\label{Conc}
We presented an approximate markov-chain based model for analyzing the dynamics of finite-buffer line networks. The model provides an iterative procedure for computing the distribution of buffer occupancy as a step in the estimation of throughput capacity of such networks. The model was seen to provide an accurate computation of throughput for varying buffer sizes, hop-length and channel erasure probabilities. The computed buffer occupancy distribution was then used to study the distribution of packet delay in such networks. This proposed model was used to identify the level of congestion in intermediate nodes, yielding a rule for intelligent memory allocation in such networks. The proposed scheme was seen to near-precisely track the dynamics and variations of the investigated performance metrics.
\vspace{-3mm}
\bibliographystyle{ieeetr}
\bibliography{ISITRef}
\end{document}